# Transition of thermal rectification in silicon nanocones

Zhongwei Zhang[a], Yuanping Chen[a*], Yuee Xie[a], and Shengbai Zhang[b†]

[a]Department of Physics, Xiangtan University, Xiangtan 411105, Hunan, P.R. China

[b]Department of Mechanical, Aerospace and Nuclear Engineering, Rensselaer Polytechnic Institute, Troy, NY, 12180, USA

**ABSTRACT:** Current understanding of thermal rectification asserts that the rectification ratio ($R$), which measures the relative heat flux between two ends of a nanostructure, is determined by its geometric asymmetry. The higher the asymmetry, the higher the $R$. However, by using nonequilibrium molecular dynamics method we have calculated thermal transport in Si nanocones as an example, the results show that such an understanding may be incorrect and $R$ may not increase monotonically with geometric asymmetry. Rather, $R$ exhibits a sharp reverse when the vertex angle ($\theta$) of the nanocone is approximately 90°. In other words, when $\theta > 90°$, R decreases, rather than increasing. We show that this abnormal behavior is originated from a change in the thermal transport mechanism. At small $\theta s$, phonon transport is dominated by localized modes, especially for transport from tip to bottom. At large $\theta s$, however, these localized modes disappear, leading to $R$ decrease.

**Keywords:** thermal rectification; transition; degree of asymmetry; Si nanocones

## 1. Introduction

Phononics requires accurate control of phonons, just like the control of electrons [1-8]. Thermal rectification is an example of controlling phonon transport in a preferred direction. Based on thermal rectification phenomena, phononic devices like thermal transistors, thermal logic circuits and thermal diodes can be developed and utilized in nanoelectronic cooling as well as thermal memory and computations [5-7]. These great potential applications have made thermal rectification attracted considerable attentions [9-12].

Thermal rectification was first found at the interface of two bulk materials [13,14]. Different thermal conductivities of the two materials and thermal potential barrier at the interfaces are considered as the origination of rectification. In nanostructures, more factors can lead to rectification phenomena because their thermal properties are more sensitive to the external modifications. Therefore, not only two nanostructures can generate rectification by interface effect [9,15-19], but also one nanostructure can also generate rectification by non-uniform mass-loading, asymmetric geometry, *et al.* [9,20-23].

Recently, thermal rectification in nanostructures with asymmetric geometry has been a focus of study [22-25], because the recent advanced technology can fabricate various asymmetric structures easily. For example, it has been shown that, in a triangle graphene, there exists a thermal rectification. Moreover, the larger the graded geometric asymmetry, the larger the rectification ratio ($R$) [22,26]. Thermal rectification has been found in a three-dimensional (3D) pyramidal diamond, and a larger geometric asymmetry would increase $R$ [24]. An asymmetric three-terminal graphene nanojunctions we studied before also exhibits a strong thermal rectification, and the structures with larger geometric asymmetry has a higher thermal rectification [23]. These studies suggest the importance of asymmetry to rectification. In these asymmetric structures, thermal rectification is usually attributed to the different temperature-dependence of the thermal conductivities of two sides and the different phonon spectra mismatch before and after reversing the applied temperature bias [24,27-29]. So, it seems that the higher the asymmetry, the higher the $R$.

However, in this paper, we find that the relation between the asymmetric degree and the $R$ is not the only case, by studying thermal transport in a three-dimensional (3D) Si nanocones with non-equilibrium molecular dynamics (NEMD) simulations. We find that $R$ of the nanocones does not increase monotonously with the geometric asymmetry. A sharp transition takes place at the vertex angle $\theta \approx 90°$. Before this point, $R$ increases with the degree of asymmetry, similar to what has been reported in the literature. However, $R$ decreases as $\theta > 90°$, i.e., when the asymmetry becomes more drastic. Our analysis suggests that the transition is a result of having two competing thermal transport mechanisms: phonon transport in small-$\theta$ nanocones is dominated by localized modes, especially for the transport from nanocone tip to bottom. With an increase in $\theta$, however, these localized modes gradually evolve into delocalized modes. Because the evolution speed of modes at the bottom is noticeably different from those at the tip, an exotic transition in $R$ takes place.

## 2. Method and model

NEMD simulations are performed to obtain the thermal properties of Si nanocones by using the LAMMPS package [30]. Stillinger-Weber-type potential with both two- and three-body terms is adopted for the interaction between Si atoms [31,32]. The time step is 0.5 fs. To reach the non-equilibrium steady state, the system was relaxed for 60 ps under NVE ensemble. After 50 ps, the temperature difference has reached the steady state. During this procedure, the total energy and average temperature both were fluctuating around targeted values. After then, 80 ps is used to calculate the results. The heat fluxes and thermal rectifications are calculated from at least four samples with different initial velocities and then obtain an averaged value for each calculation.

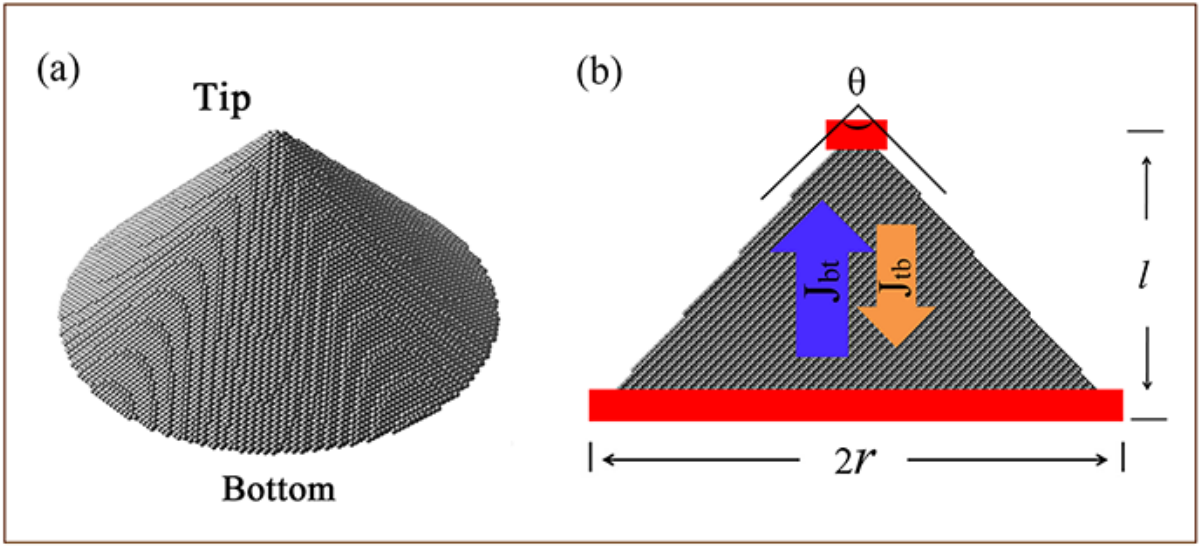

**FIG. 1.** Schematic drawing of the Si nanocones with two different viewing perspectives. (a) The 3D perspective with a round bottom and pointed tip. (b) The 2D vertical cross section with structural parameters. Here, $r$ is the radius of the bottom, $l$ is the length in the vertical direction and $\theta$ is the angle at the tip. The (red) ends denote contacting heat bathes.

Figure 1 shows Si nanocones, where $l$, $r$, and $\theta$ define its height, radius at the bottom, and vertex angle, respectively. There are two reasons to consider such structures: first, these 3D nanocones possess large geometric asymmetry because of the large area ratio between the bottom and tip of the cones. Second, these nanocones are a special type of Si quantum dots, which can be readily fabricated by epitaxial growth on Si(100) surface [33,34]. In our study, the tip and bottom of the nanocone are put in contact with two heat baths at $T_t=T_0(1-\Delta)$ and $T_b=T_0(1+\Delta)$, respectively, where $T_0$ is the average temperature and $\Delta$ is the normalized temperature bias. The outside layers of the heat baths are fixed during the calculation to avoid spurious global rotation. For $\Delta > 0$, a heat flux $J_{bt}$ will flow from bottom to tip [see Fig. 1(b)]. For $\Delta < 0$, on the other hand, are verse flux $J_{tb}$ will flow from tip to bottom. If $J_{tb} \neq J_{bt}$, then thermal transport across the nanocone is asymmetric. It means thermal rectification with:

$$R = \frac{J_{bt} - J_{tb}}{J_{tb}} \times 100\%. \tag{1}$$

## 3. Results and discussion

To understand the relationship between geometric asymmetry and thermal rectification, thermal transport in nanocones of different $\theta$ is studied. The inset in Fig. 2(a) shows three side views for nanocones with $\theta = 45, 85$, and $120°$, respectively, where the length of the nanocone is fixed at $l = 3.26$ nm. The radius of the bottom is increased from $r = 1.35$ to $5.65$ nm, as $\theta$ is increased from $45°$ to $120°$. Figure 2(a) shows $J_{tb}$ and $J_{bt}$ as a function of $\theta$ for $T_0 = 300$ K and $|\Delta|$ = 0.3. It shows that the geometric asymmetry always induces a larger $J_{bt}$ than its inverse $J_{tb}$.

Hence, as expected, phonons prefer to transport from a wide bottom to a sharp tip, but not the other way around. What is unexpected, however, is that $J_{tb}$ and $J_{bt}$ can exhibit a very different dependence on $\theta$, with $\theta = 90°$ being the inflection point. For $\theta < 90°$, $J_{bt}$ increases with $\theta$ considerably faster than $J_{tb}$. For $\theta > 90°$, on the other hand, both $J$s increase fast with $\theta$. This change leads to a qualitative change in $R$ according to Eq. (1).

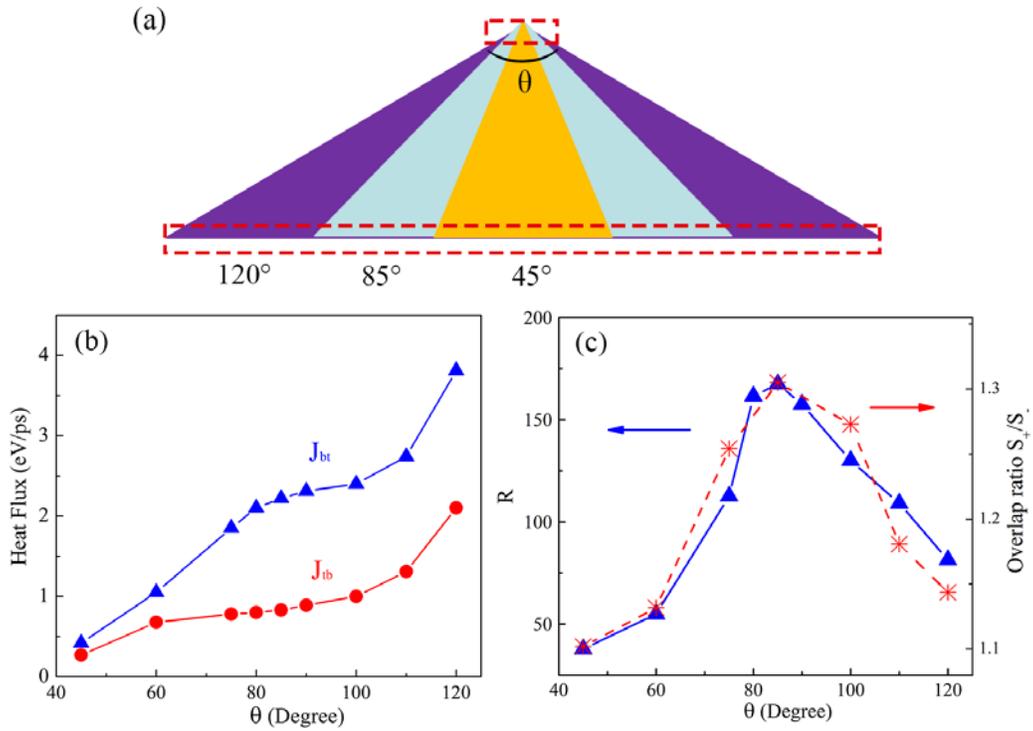

**FIG. 2.** (a) Cross sections of nanocones with vertex angle $\theta$ = 45, 85 and 120°, respectively. In the plot, the vertical length $l$ is fixed. Open bars (enclosed by dashed red lines) denote contacting heat bathes. (b) Heat fluxes $J_{bt}$ and $J_{tb}$ as function of $\theta$. (c) Rectification ratio (left vertical axis) and normalized $S_+/S_-$ (right vertical axis) as function of $\theta$. In (b) and (c), $l$ = 3.26 nm, $T_0$ = 300 K, and $|\Delta|$ = 0.3.

Figure 2(c) shows the relationship between $R$ and $\theta$ (see solid line). For $\theta < 90°$, $R$ increases with θ, because $J_{bt}$ increases faster than $J_{tb}$. For θ > 90°, R decreases rather than increases with θ because of a large $J_{tb}$. These results indicate that in Si nanocones *R* does not increase monotonously with geometric asymmetry, which is at variance with the conventional theory.

In order to determine the physical origin behind the unexpected results in Fig. 2, we first analyze the vibrational characteristics of the nanocones at different temperature biases (Δ > 0 and Δ < 0). The vibration density of states (*vDOS*) in the nanocone are nonuniform because of the asymmetric shape. One cannot find significant message if just compare the *vDOS* of the whole structure for opposite Δ. Therefore, we compare the *vDOS* of tip and bottom regions after switching temperature, which is a common method used to explain thermal rectification in two-segment structures and also in graded structures. The regional v*DOS* can be calculated from the Fourier transform of the velocity autocorrelation function [27,29,35]:

$$\text{vDOS}(\omega) = \frac{1}{\sqrt{2\pi}} \int e^{i\omega t} \langle \sum_{j=1}^{N} v_j(t) v_j(0) \rangle \, dt, \qquad (2)$$

where $v_j$ is the velocity vector for particle *j*, ω is the vibration wavenumber, and *N* is the number of atoms in the region of interest. For the two sides of a nanocone, v*DOS* should be different because the temperature difference and geometric asymmetry. The overlap,

$$S = \frac{\int_0^\infty \text{vDOS}_b(\omega) \text{vDOS}_t(\omega) d\omega}{\int_0^\infty \text{vDOS}_b(\omega) d\omega \int_0^\infty \text{vDOS}_t(\omega) d\omega}, \qquad (3)$$

of *vDOS* between the two sides provides an indirect measure of the phonon transmission spectra. A larger *S* usually suggests a larger heat flux. Therefore, the overlap ratio $S_+/S_-$ is used to characterize the rectification properties [7,29,36], where the subscripts (+) and (-) correspond to $\Delta > 0$ and $\Delta < 0$.

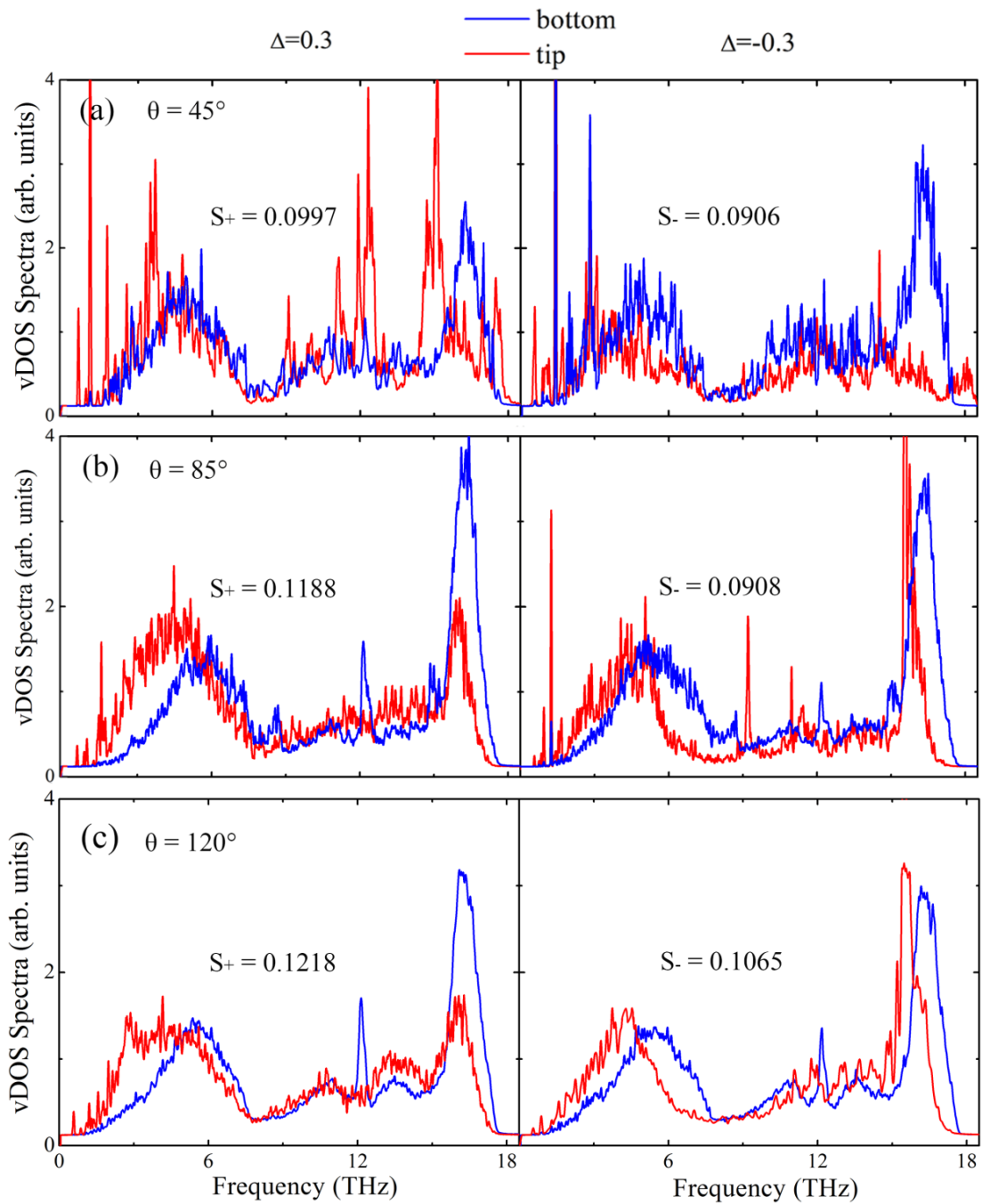

**Fig. 3.** Vibrational density of states (*vDOS*) in the bottom (blue line) and tip (red line) regions, at two different temperature gradients, $\Delta = \pm 0.3$. (a) $\theta = 45°$, (b) $\theta = 85°$, and (c) $\theta = 120°$. The corresponding spectral overlaps $S_{\pm}$ are indicated. Same as in Fig. 2, $l = 3.26$ nm and $T_0 = 300$ K.

Figure 3 shows, for $T_0 = 300$ K and $|\Delta| = 0.3$, the *vDOS* for tip and bottom at $\theta = 45, 85,$ and $120°$, respectively, along with calculated $S_+$ and $S_-$. The *vDOS* profiles of thex three structures are considerably different.

For the 45° nanocone in Fig. 3(a), numerous sharp and irregular peaks were observed on the spectra. Such peaks are usually caused by phonon localization on the surfaces. As the localized phonon modes occupy both the tip (red line) and bottom (blue line) sides of the 45° nanocone, especially on the tip side, they dominate the phonon transport. This results in a weak phonon transmission regardless $\Delta > 0$ or $< 0$, which is supported by the calculated small $S_+ = 0.0997$ and $S_- = 0.0906$. Accordingly, both $J_{tb}$ and $J_{bt}$ are small, so is $R$.

For the 85° nanocone in Fig. 3(b), on the other hand, the spectrum of the bottom side becomes smoother while that of the tip side still has a number of sharp and irregular peaks. This implies that, in the 85° nanocone, phonon localization at the bottom side has faded out, while at the tip side they are still somewhat localized. Meanwhile, one finds that the spectrum at the tip side is rather sensitive to the temperature, by comparing the red lines on the left and right panels of Fig. 3(b). For most frequencies, v*DOS* at $\Delta = 0.3$ is larger than that at $\Delta = -0.3$, especially in the low-frequency regime. Such a difference leads to a large difference between $S_+$ and $S_-$, between $J_{bt}$ and $J_{tb}$, and subsequently a large $R$. Hence, large $R$ is a result of having different phonon modes at the two sides of the nanocone.

For the 120° nanocone in Fig. 3(c), the spectra become smooth on both side. The difference in v*DOS* between the two sides becomes small. Hence, for the 120° nanocone, the delocalized phonon transport is taken over by delocalized phonon transport of the propagation modes. In this case, both $S_{+/-}$ and $J_{bt/tb}$ are large to result in a small $R$.

The above results indicate that the evolution of the phonon modes on the two sides of the nanocone with geometry is qualitatively different. With increasing $\theta$, the change of the phonon modes from localized to delocalized is faster on the bottom side than on the tip side. One can quantify the different changes of phonon modes by plotting the overlap ratio $S_+/S_-$ as a function of $\theta$, shown in Fig. 2(c). The ratio coincides with the variation of $R$, confirming that the unconventional change in $R$ is caused by the out-of-phase change in the phonon localization. Furthermore, the evolution speed of phonon modes is related to the increase of atoms numbers or volume on the two sides [24], which is reflected in the disappearance of sharp peaks and appearance of new peaks of vDOS. The increase of atoms numbers leads to the phonon modes of two sides changing from localized to delocalized, while the faster increase on the bottom side results in faster evolution of phonon modes.

One may also examine the phonon localization from a different perspective by defining the phonon participation ratio [28,37,38]:

$$p_\lambda^{-1} = N \sum_i \left( \sum_\alpha \varepsilon_{i\alpha,\lambda}^* \varepsilon_{i\alpha,\lambda} \right)^2, \qquad (4)$$

where *i* runs over all *N* atoms, $\varepsilon_{i\alpha,\lambda}$ for a given phonon mode $\lambda$, is the *α*-component of the eigenvalue on the *i*th atom, and $\alpha$ = X, Y, and Z is the polarization. The calculated $p_\lambda$ is shown in Fig. 4. As expected, the nanocones generally have a smaller $p_\lambda$ than bulk, due to the presence of localized modes. Also due to increased delocalization, $p_\lambda$ monotonically increases with $\theta$: for example, its mean value in Figure 4 is $p_\lambda$ = 0.5 for 45° nanocone, 0.6 for 85° nanocone, and 0.7 for 120° nanocones, respectively. These results support the qualitative discussion of phonon localization modes in Figure 3.

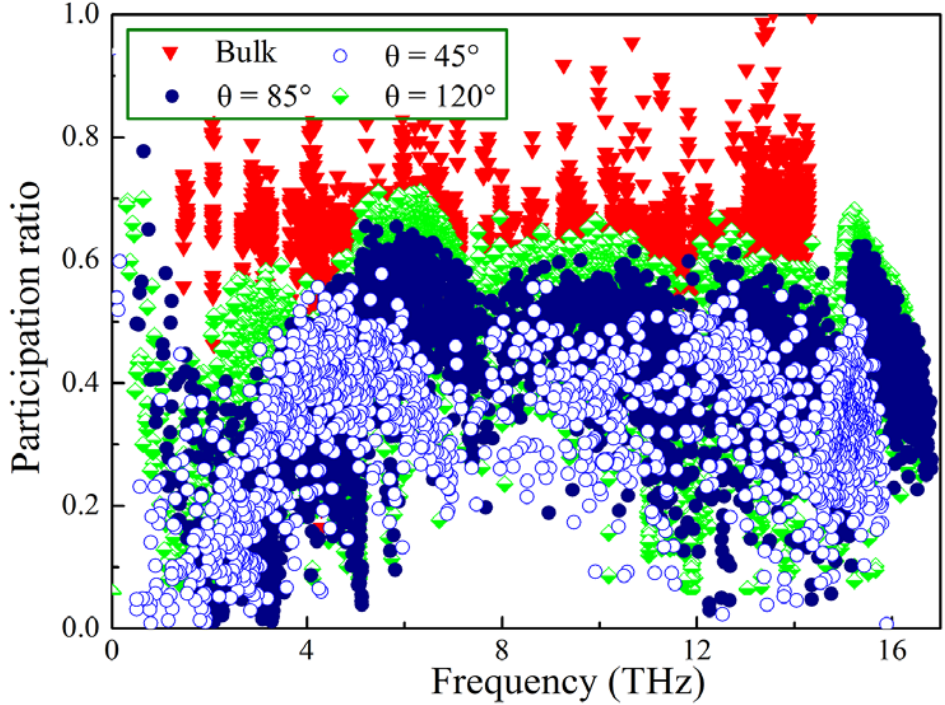

**FIG. 4.** Participation ratio $p_\lambda$ (defined in the text) of Si bulk and nanocones with $\theta$ = 45, 85, and 120°, respectively, as functions of phonon frequency.

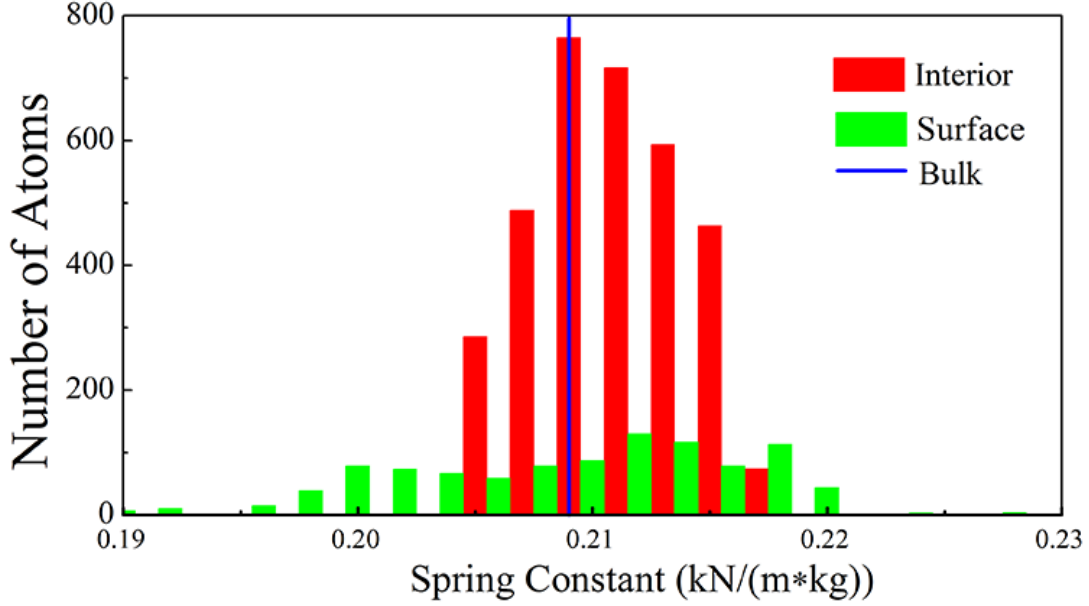

**FIG. 5.** Number of surface and interior atoms for $\theta = 85°$ nanocone, as a function of spring constant. $l = 3.26$ nm. The value for bulk Si is also indicated.

The phonon localization is also reflected in the surface-to-volume ratio (SVR) [39,40]. This is because surface atoms are placed in a different force environment than bulk atoms with fewer neighboring atoms. Therefore, the spring constants of the surface atoms and interior atoms are noticeably different. Figure 5 shows the spring constants for the 85° nanocone. It reveals that the spring constants for surface atoms are considerably wider than those for interior atoms. This difference in the force constants is the reason for localized phonon modes. In the nanocones, SVRs are inversely proportional to $\sin\frac{\theta}{2}$. Hence, the degree of phonon localization decreases with increasing $\theta$. In addition, SVRs for the two sides of the nanocone are also different. The SVRs of tip are approximately inversely proportional to $\sin\frac{\theta}{2}$, while those of bottom are approximately inversely proportional to $\tan\frac{\theta}{2}$. Therefore, the evolutions of the localized modes on the two sides of the nanocone with increasing $\theta$ are out of phase (as discussed above).

To further understand the evolution of the phonon modes, we may visualize the energy spatial distribution of phonons in the overall structure, which is defined as [28,38,39]:

$$E_i = \sum_\omega \left(n + \frac{1}{2}\right) \hbar \omega D_i(\omega), \tag{5}$$

where $n$ is the occupation number from Bose-Einstein distribution, and $D_i(\omega)$ is the local vibrational density of states (*lvDOS*): $D_i(\omega) = \sum_\lambda \sum_\alpha \varepsilon^*_{i\alpha,\lambda} \varepsilon_{i\alpha,\lambda} \delta(\omega - \omega_\lambda)$.

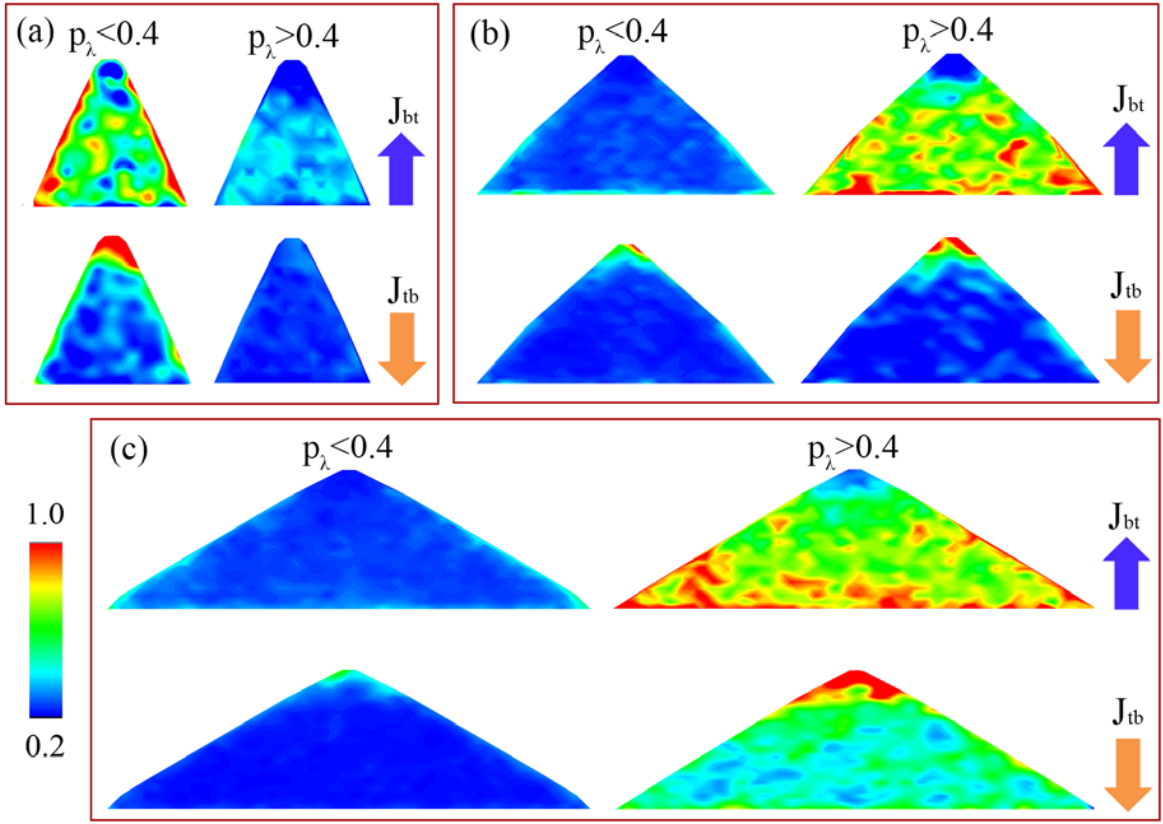

**FIG. 6.** Energy spatial distribution for the two heat flow directions ($J_{bt}$ and $J_{tb}$) at $0 \leq p_\lambda \leq 0.4$ and $0.4 \leq p_\lambda \leq 1$, respectively. (a) $\theta = 45°$, (b) $\theta = 85°$, and (c) $\theta = 120°$. $l = 3.26$ nm, $T_0 = 300\ K$, and $|\Delta| = 0.3$.

Figure 6 shows $E_i$ for 45, 85, and 120° nanocones at $T_0$ = 300 K and $|\Delta|$ = 0.3. Each energy spatial distribution is divided into two parts, one is the distribution of localized modes, i.e., sum of $\omega$ in Eq. (4) for $0 < p_\lambda < 0.4$, the other is the distribution of delocalized modes, i.e., sum of $\omega$ for $0.4 < p_\lambda < 1.0$. For the 45° nanocone, almost all phonon energies are gathered in the localized modes, as shown in Fig. 6(a). For $J_{bt}$, most phonons are localized on the surface of the sidewall of the nanocone. For $J_{tb}$, on the other hand, phonons are tightly localized at the tip of the nanocone. These localized phonon modes dominate the transport and thus the heat fluxes in the nanocones are small regardless the flux direction. However, there are still important difference between $J_{bt}$ and $J_{tb}$: for $J_{tb}$, the localized phonons around the tip lead to a bottleneck effect in the transport from tip to bottom [28]. For the 85° nanocone, the difference of $E_i$ in the tip and bottom becomes more pronounced (see Fig. 6(b)). For $J_{bt}$, the localized modes are only distributed on the surface of the side wall. The delocalized modes are, on the other hand, distributed throughout the nanocone. This implies a transport of large heat flux $J_{bt}$ from bottom to tip. For $J_{tb}$, the bottleneck effect caused by the localized modes also exists, and hence the heat flux $J_{tb}$ remains small. The large difference between $J_{tb}$ and $J_{bt}$ leads to a large $R$. For the 120° nanocone, Fig. 6(c) shows that most of the phonon modes are delocalized and phonon energies are distributed through the nanocone regardless the temperatures at the two sides. In this case, phonon transport is dominated by the propagation modes, which decreases $R$ by increasing the denominator in Eq. (1). From these discussions, one can further conclude that the evolution speeds of phonon modes on the two sides are different.

In addition, the transition can also be interpret from another point of view. We can define a ratio of volumes $r_V = V_t/V_b$, where $V_b$ and $V_t$ represent the volumes of bottom and tip sides, respectively. $r_V$ is a constant approximately when $\theta$ is changed. At the small angle $\theta \sim 0°$, both $V_b$ and $V_t$ are small. In this case the nanocone is somewhat similar to a nanowire because the sizes of the bottom and tip sides are small, and thus phonon localizations can be found in both sides. At the angle $\theta \sim 180°$, however, both $V_b$ and $V_t$ become very large, i.e., the nanocone is similar to a bulk, in which the localized modes evolve into bulk phonon modes. From this point of view, the two limiting cases are of small geometric asymmetry, while the maximum geometric asymmetry is at $\theta \sim 90°$. Therefore, the transition of R occurs.

## 4. CONCLUSIONS

In conclusion, we have studied the thermal rectification properties of 3D Si nanocones by using molecular dynamics simulations. The results show that the $R$ of the nanocones does not increase monotonically with geometric asymmetry. A transition of $R$ appears at $\theta = 90°$: below this angle, $R$ increases with geometric asymmetry, while above this angle, $R$ decreases with geometric asymmetry. The $vDOS$ spectra, phonon participation ratio, and spatial distribution energy are all in support of the calculated unconventional transition and reveal that the transition is caused by the different evolution speed of phonon modes at the two sides of the nanocones. Finally, we hope some experiments can be carried out to validate and extend our findings.

**Acknowledgments**

We thank discussion with T. M. Lu. This work was supported by the National Natural Science Foundation of China (Nos.51176161, 51376005 and 11474243). SBZ was supported by the US National Science Foundation DMR-1305293.

**Corresponding author:** * chenyp@xtu.edu.cn; † zhangs9@rpi.edu.